\documentclass[12pt]{article}
\usepackage[dvips]{graphicx}
\begin{document}
\title{Nonlinear structure in a current-carrying collisional dusty plasma}
\author{F. Haas \footnote{Also at Universidade do Vale do Rio dos Sinos - UNISINOS, Av. Unisinos 950, 93022--000, S\~ao Leopoldo, RS, Brazil} \,and P. K. Shukla}
\date{\relax}
\maketitle
\begin{center}
{Institut f\"ur Theoretische Physik IV, 
Ruhr-Universit\"at Bochum\\ D-44780 Bochum Germany}
\end{center}

\begin{abstract}
A perpendicular ion drift is investigated as a mechanism for the generation of magnetic field structures in a highly collisional dusty plasma. The basic dissipation mechanism is assumed to be the dust-neutrals momentum exchange, so that plasmas with a small ionization fraction are natural candidates for experiments. The model reduces to a nonlinear partial differential equation for the vector potential. The conditions for linear instability are determined. The stationary states are periodic arrangements for the magnetic field, described by a Lienard equation. The fully depleted (ion-dust) case is also considered in detail. Applications to planetary rings, comets  and dust acoustic instability experiments are analyzed. A necessary condition for the validity of the model is a sufficiently slow time-scale of the generated magnetic fields. 
\end{abstract}

\section{Introduction}
Collisional processes have an important r\^ole in dusty plasmas in laboratory or in the environment, as for the dust-acoustic and dust-ion-acoustic wave propagation and instabilities \cite{Yaroshenko, Ivlev}. In particular \cite{Shukla1}, collisions between the charged dust particles and the neutrals are relevant, in view of the usually small degree of ionization in applications. Ion-neutrals collisions are also fundamental in experimental observation of dust-acoustic waves \cite{Barkan}, while dust-neutrals collisions are related to the generation of dust-acoustic waves driven by gravity \cite{DA}. On the other hand, drifting populations are ubiquitous in complex plasmas. This occurs for ion-dust streaming instabilities \cite{Rosenberg}, electrostatic dust-cyclotron instabilities in collisional magnetoplasmas \cite{DAngelo} and for current-driven ion-acoustic instabilities in complex plasmas \cite{Merlino2, Merlino}. 

In most of the above applications \cite{Yaroshenko, Rosenberg2}, there is some streaming between electrons and/or ions relative to the dust, in the same direction of the motion. In contrast, the present work consider a {\it perpendicular} drift of the ions, relative to the bulk dust motion. 
Ions and electrons are taken in local thermodynamical equilibrium, which is equivalent to disregarding their inertia, while the dust fluid is assumed to be cold. Similar models were proposed for dust kinetic Alfv\'en waves in dust-ion plasmas \cite{Mizra}, but without the perpendicular ion drift. 
As shown in the following, this ion drift is responsible for the generation of periodic magnetic field patterns. Notice that structures like braids, kinks and filaments at Jupiter's rings can be a consequence of an ambient magnetic field pattern and the associated force balance for the charged dust particles \cite{Avinash}. 

This work is organized as follows. The necessary equations describing the electron, ion and dust fluids and the  electromagnetic field are proposed in Section 2. Assuming a slab geometry, the system is then reduced to a nonlinear partial differential equation for the $y$-component of the vector potential. This equation is studied in the linear and in the stationary limit. Linear solutions are shown to be unstable and to present a singular growth-rate for small wave-numbers.  In addition, it is proved in the Appendix that the divergence of the growth-rate is related to a too fast characteristic time-scale, outside the range of validity of the model. On the other hand, the stationary solutions are described by a nonlinear second-order ordinary differential equation, the Lienard equation, whose solutions shown periodic behavior.  In Section III, the illustrative and relevant situation where electrons are fully depleted into the dust grains is discussed. Here the fundamental partial differential equation for the vector potential reduces to a Burger's like equation and there is no singularity in the dispersion relation. Section 4 is dedicated to the final remarks. 

\section{Basic equations}

Consider a plasma composed by electrons and ions, as well as by massive, negatively charged dust grains. As a first approximation, we take inertialess (in comparison to the dust particles) electrons and ions. For characteristic time scales much greater than the typical dust charge fluctuation time, the dust charge can be taken as a constant. Assuming a slab geometry, the basic equations are 
\begin{eqnarray}
\label{e1}
\frac{\partial n_d}{\partial t} &+& \frac{\partial}{\partial x} (n_{d} v_{dx}) = 0 \,, \\
\label{e2}
0 &=& - Z_d e E_x - m_d \nu_{dn} v_{dx} \,, \\
\label{e3}
0 &=& e (E_x - v_0 B_z )- \frac{\kappa_B T_i}{n_i} \frac{\partial n_i}{\partial x} \,,\\
\label{e4}
0 &=& - e E_x - \frac{\kappa_B T_e}{n_e} \frac{\partial n_e}{\partial x} \,,\\
\label{e5}
\frac{\partial B_z}{\partial x} &=&  \mu_0 e (n_i - n_{i0} ) v_0   \,, \\
\label{e6}
Z_d n_d &=& n_i - n_e \,.
\end{eqnarray}
In (\ref{e1}-\ref{e6}), the ion, electron and dust charges are $e$, $-e$ and $q_d = -Z_d e$ respectively, where $Z_d$ is the dust charge number and $e$ the elementary charge. In addition, $n_{i}, n_{e}$ and $n_d$ are the ion, electron and dust particle densities. The ions are assumed to have a perpendicular drift velocity $- v_0 \hat{y}$, the negative sign chosen for later convenience. Drifting electrons could also be present, but here ions were preferred since in laboratory and astrophysical plasmas frequently the electron population is negligible ($n_e \ll n_i$) due to electron charge depletion into dust grains \cite{Barkan}. Also, any current arising from dust grain flows is neglected, since it is smaller than the ion current by a factor $(m_{i}/m_{d})^{1/2}$ at least, where $m_i$ and $m_d$ are the ion and dust masses, respectively. In this context, the magnetic field generation is due to the drifting ions only.  Also, in the above equations ${\bf B} = B_{z}(x,t) \hat{z}$ is the magnetic field, $E_x$ the $x$ component of the electric field, $v_{dx}$ the $x$ component of the dust velocity, $\nu_{dn}$ the dust-neutrals collision frequency and $T_{i,e}$ the ion (electron) temperature, while the dust fluid is cold. Collisions between electrons and neutrals or between ions and neutrals are not taken into account. From Amp\`ere's law (\ref{e5}) it is implicit that $n_i$ is the total and $n_{i0}$ is the average ion number densities, so that $\partial B_{z}/\partial x \rightarrow 0$ as $n_i \rightarrow n_{i0}$. A similar model \cite{Niknam} was recently proposed for a current-carrying electron-ion, non-dusty plasma. 

In (\ref{e6}), quasineutrality is assumed, which is a good approximation for low-frequency phenomena. For the same reason, the displacement current is neglected, as well as the additional drifts arising from the generated magnetic field.  In this context, disregarding these Hall-type currents is the most delicate assumption, more throughly discussed in the Appendix. There it is shown that these perpendicular motions can be discarded in a sufficiently slow time-scale. The ion drag force due to momentum transfer between ions and charged dust grains is also not taken into account. This imply, in particular, that the treatment is adequate for the co-rotating distance at the main $(A-C)$ Saturn's rings, so that $|v_{i0}-v_{d0}| < v_{Ti}$,  where $v_{i0}$ and $v_{d0}$ are the equilibrium ion and dust speeds and $v_{Ti}$ is the thermal speed of ions. For outer rings, the plasma ions overtake the dust grains, while for smaller distances the ions are slower than the dust fluid. In both cases, significant momentum transfer takes place and more detailed models are needed \cite{Yaroshenko}. However, the present work is basically concerned with the emergence of magnetic field patterns in highly collisional dusty plasmas, so that these effects are neglected.  The model is still valid for the co-rotating distance at Saturn's rings (where $x$ refers to the azimuthal direction) and for specially devised laboratory experiments. 

From now on we use the rescaled variables 
\begin{eqnarray}
\label{e11}
\bar{n}_e &=& n_{e}/n_{i0} \,, \quad  \bar{n}_i =  n_{i}/n_{i0} \,, \quad  \bar{n}_d =  Z_{d} n_{d}/n_{i0} \,, \\
\label{e12}
\bar{E}_x &=& eE_{x}/(\kappa_{B}T_i k_{0}) \,, \quad \quad \bar{B}_z = e\lambda_{Di}c B_{z}/(\kappa_{B}T_{i}) \,,\\
\label{e13}
\bar{x} &=& k_0 x \,, \quad  \bar{t} = \omega_{0}t \,, \quad \bar{v}_{dx} = k_0 v_{dx}/\omega_0 \,,
\end{eqnarray}
for 
\begin{equation}
\label{k0}
k_0 = \frac{v_{0}}{\lambda_{Di} c} \,, \quad \omega_0 = \frac{Z_d m_i  \omega_{pi}^2 v_{0}^2}{m_d \nu_{dn}c^2} \,,
\end{equation}
where  $c$ is the speed of light, $\lambda_{Di}  = (\epsilon_{0}\kappa_{B}T_{i}/(n_{i0} e^{2}))^{1/2}$ the ion Debye length and $\omega_{pi} = (n_{i0}e^{2}/(m_{i}\epsilon_{0})^{1/2}$ the ion plasma frequency. As will be seen in the continuation, $k_0$ plays a major r\^ole as a maximal wave-number for linearly unstable modes. Also, it is implicit from (\ref{e13}) that all wave-numbers are normalized to $k_0$ and that all frequencies and growth-rates are normalized to $\omega_0$. 

For simplicity of notation, the bars will be dropped Then, if the electrostatic and vector potentials are denoted by $\phi = \phi(x,t)$ and ${\bf A} = A_{y}(x,t)  \hat{y}$, so that $E_x = - \partial\phi/\partial x$, $B_z = \partial A_{y}/\partial x$, then (\ref{e2}-\ref{e4}) and (\ref{e6})  gives
\begin{eqnarray}
\label{e7}
n_e &=& \rho \exp(T_{i}\phi/T_{e}) \,,\\
\label{e8}
n_i &=& \exp(-\phi - A_{y}) \,,\\
\label{e9}
n_d &=& \exp(-\phi - A_{y}) - \rho \exp(T_{i}\phi/T_{e})  \,,\\
\label{e10}
v_{dx} &=& \partial\phi/\partial x \,,
\end{eqnarray}
where $\rho = n_{e0}/n_{i0}$ is the ratio between the equilibrium electron and ion densities. Notice that the $y$ or $z$ components of the model equations as well as Faraday's law are of no concern, since (\ref{e1}-\ref{e6}) form a closed system. 

Eqs.  (\ref{e5}) and (\ref{e8}) allows to express
\begin{equation}
\label{e14}
\phi = - A_y - \ln(1 + \frac{\partial^2 A_y}{\partial x^2}) \,,
\end{equation}
while then (\ref{e9}) gives 
\begin{equation}
\label{e15}
n_d = 1 + \frac{\partial^2 A_y}{\partial x^2} - \rho\exp\left(-\frac{T_i}{T_e}(A_y + \ln(1 + \frac{\partial^2 A_y}{\partial x^2}))\right) \,.
\end{equation}
Taking into account (\ref{e10}-\ref{e15}), all quantities can be written in terms of $A_y$ alone. The only remaining equation is (\ref{e1}), which leads to 
\begin{eqnarray}
\frac{\partial}{\partial t}\Biggl[\frac{\partial^2 A_y}{\partial x^2} &-& \rho\exp\left(-\frac{T_i}{T_e}(A_y + \ln(1 + \frac{\partial^2 A_y}{\partial x^2}))\right)\Biggr] =  \nonumber  \\ 
\label{e16}
&=& \frac{\partial^2}{\partial x^2}\Biggl[\frac{\partial^2 A_y}{\partial x^2}  + \frac{1}{2}\left(\frac{\partial A_y}{\partial x}\right)^2 + A_y + \\ &+&
\frac{\rho T_e}{T_i} \exp\left(-\frac{T_i}{T_e}(A_y + \ln(1 + \frac{\partial^2 A_y}{\partial x^2}))\right) \Biggr] \,. \nonumber
\end{eqnarray}

The resulting equation seems to be too complicated to be treated in full generality. In this Section, only the linear and the stationary limits will be considered. In Section III, the complete depletion case ($\rho  \equiv 0$) will also be shown to be amenable to some more detailed analysis. 

In the linear regime, (\ref{e16}) reduces to 
\begin{equation}
\label{e17}
\frac{\partial}{\partial t}\left[\frac{\partial^2 A_y}{\partial x^2} + \frac{\rho T_i}{T_e} \left(\frac{\partial^2 A_{y}}{\partial x^2} + A_y \right)\right] = (1 - \rho ) \frac{\partial^2}{\partial x^2} \left[\frac{\partial^2 A_y}{\partial x^2} + A_y \right] \,.
\end{equation}
Assuming $A_y \sim \exp[i(kx-\omega t)]$ then gives the dispersion relation
\begin{equation}
\label{e18}
\omega = \frac{i\,(1-\rho )(1-k^2)k^2}{k^2 - (\rho T_{i}/T_{e}) (1-k^2)} \,.
\end{equation}
Taking into account that $\rho  < 1$ because the dust grains are negatively charged, (\ref{e18}) corresponds to instability ($Im(\omega) > 0$) provided
\begin{equation}
\label{e19}
k_{-}^2 < k^2 < 1 \,,  \quad {\rm where} \quad k_{-} = \left(1 + \frac{T_e}{\rho T_i}\right)^{-1/2} \,,
\end{equation}
otherwise there is an evanescent mode. In addition, in terms of dimensional variables, (\ref{e19}) reveals that the wave-number $k_0$ corresponds to marginal stability, with $k_0$ given by (\ref{k0}).  Also, for $\rho  \rightarrow 1$, so that $n_d \rightarrow 0$, the growth-rate becomes smaller. In other words, the present instability is intrinsically dependent on the dust, disappearing for electron-ion plasmas. 

For $k \rightarrow  k_{-}$, the growth-rate becomes singular. In several applications, one has $k_{-}$ as a number not too far from unity. This corresponds to wavelengths which are small but still much higher than the ion Debye wavelength. Therefore, a breakdown of the quasineutrality assumption is not a sufficient reason for ruling out the singularity as unimportant. Moreover, if the Poisson equation is taken into account it can be shown that a (slightly modified) singularity is still present. Also notice that the singularity disappears when $\rho  \rightarrow 0$. 

The clue to solve the problem is to notice that, when the growth-rate is too large, all quantities, and the magnetic field in particular, would change in a fast time-scale, in contradiction to the hypothesis of the model. This is a direct consequence of the neglect of the additional drifts in the $y-$direction due to the generated magnetic field (Hall effect). To prevent the emergence of a fast time-scale  behavior, the wave-number has to be further bounded from bellow according to 
\begin{equation}
\label{km}
k > k_m > k_{-} \,,
\end{equation}
where the full expression for $k_m$ is derived in the Appendix, see (\ref{a6}).  

\begin{figure*}[ht]
\begin{center}
{\par\centering
\resizebox*{0.80\textwidth}{!}{\rotatebox{0}{\includegraphics{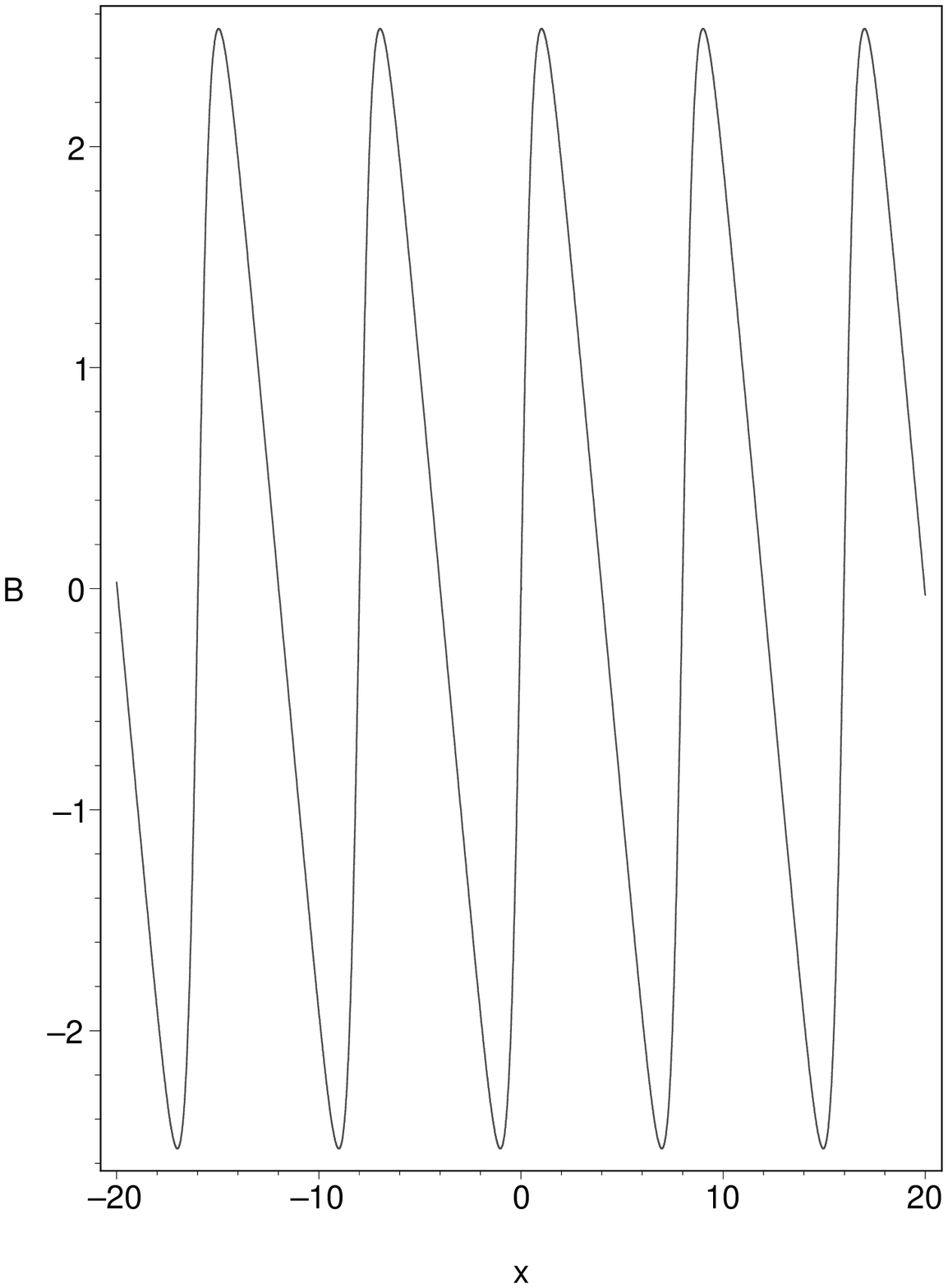}}}
\par}
\caption{$B(x)$ from (\ref{e21}) for the sample initial conditions $B(0) = 0, B'(0)=5$.}
\label{fig1}
\end{center}
\end{figure*}

Besides the small amplitude case, another relevant limit is the stationary one, so that $\partial A_{y}/\partial t \equiv 0$. In this case, the left-hand side of (\ref{e16}) identically vanishes. The right-hand side, after integrated once assuming, for the sake of definiteness, that  $dA_{y}/dx$ and $d^{3}A_{y}/dx^3$ vanishes at $x = 0$, we derive, after some simple algebra, 
\begin{eqnarray}
\label{e20}
\Biggl(\frac{d^3 A_y}{dx^3} &+& \frac{dA_y}{dx}(\frac{d^2 A_y}{dx^2} + 1)\Biggr) \times  \\
&\times& \left(1 - \rho  \frac{\exp\left[-(T_{i}/{T_e})(A_y + \ln(1 + d^2 A_{y}/{dx^2}))\right]}{1 + d^2 A_{y}/dx^2}\right) = 0 \,. \nonumber
\end{eqnarray}
Defining $B = dA_{y}/dx$, one of the equilibria is found when the first factor in (\ref{e20}) is zero,
\begin{equation}
\label{e21}
\frac{d^2 B}{dx^2} + B \frac{dB}{dx} + B = 0 \,.
\end{equation}
This is in the class of Lienard equations \cite{Polyanin} and is the same equation found for the stationary magnetic field of a collisional current-carrying (non-dusty) plasma \cite{Niknam}. Observe that, by differentiation of (\ref{e14}), the magnetic fields corresponding to (\ref{e21}) are associated to zero electric fields and  dust velocities $v_{dx}$. In addition, from (\ref{e7}) the electron density would be constant, while from (\ref{e8}-\ref{e9}) it is seen that the ion and dust populations appear modulated. The Lienard equation has an exact parametric solution \cite{Niknam, Polyanin}, but here we content ourselves with a numeric solution, as in Figure (\ref{fig1}), showing a typical periodic behavior for $B(0)=0, B'(0)=5$. 

For $\rho  \neq 0$, however, there is a second possibility for the stationary magnetic field, where the second factor in (\ref{e20}) vanishes. After differentiating, this can be shown to imply 
\begin{equation}
\label{e22}
(1 + \frac{T_i}{T_e}) \frac{d^2 B}{dx^2} + \frac{T_i}{T_e} ( B \frac{dB}{dx} + B) = 0 \,.
\end{equation}
The further rescaling
\begin{equation}
\label{e23}
\tilde{B} = (1 + \frac{T_e}{T_i})^{-1/2} \,B \,, \quad  \tilde{x} = (1 + \frac{T_e}{T_i})^{-1/2} \,x 
\end{equation}
converts (\ref{e22}) into the same Lienard equation as before,
\begin{equation}
\label{e24}
\frac{d^2 \tilde{B}}{d\bar{x}^2} + \tilde{B} \frac{d\tilde{B}}{d\tilde{x}} + \tilde{B} = 0 \,.
\end{equation}
Despite the fact that both stationary states are described by the same (up to a rescaling) equation, there remains the conclusion that there are two possible nonlinear structures, provided $\rho  \neq 0$. 

\section{The complete depletion case ($\rho  \equiv 0$)}

It is well-known that complete depletion of the electronic population is not possible, since the minimum value of $\rho^2$ is the ratio between the electron and ion masses, assuming $T_e = T_i$ and point dust particles \cite{Shukla}. However, in some cases, assuming $\rho  \equiv 0$ is a useful approximation. In such dust-ion plasmas, the wave dynamics is determined only by the inertia of the dust grains and the pressure of the ion fluid \cite{Mamun}. This is a relevant scenario, in particular, for laboratory experiments on dust-acoustic waves \cite{Barkan}, certain stages of protostellar cloud collapse \cite{Rosenberg2}, the Saturn's $F$ ring \cite{Mamun} and for Halley's comet \cite{Farid}. 

Postulating $\rho = 0$, together with integrating (\ref{e16}) once, leads to 
\begin{equation} 
\label{e25}
\frac{\partial B}{\partial t} =  \frac{\partial^2 B}{\partial x^2} + B \frac{\partial B}{\partial x}  + B  \,.
\end{equation}
This is a much more simpler diffusive Burger's like equation, in comparison with (\ref{e16}). It is formally the same as the equation for the magnetic field in a current-carrying electron-ion (non dusty) plasma \cite{Niknam}.  

The results from Section 3 about linear waves and stationary states directly translate to (\ref{e25}). For the linear waves, $\rho = 0$ in (\ref{e18}) gives $\omega = i (1 - k^2)$, which is an unstable mode for $k^2 < 1$. Here there is no singularity. Concerning the stationary states, they are described by a Lienard equation, as apparent either from (\ref{e25}) with time-independent $B$ or from (\ref{e20}) with total electron depletion. Inverting the rescaling (\ref{k0}), there follows the growth-rate 
\begin{equation}
\label{g}
\gamma = \frac{Z_d m_i \omega_{pi}^2}{m_d \nu_{dn}}\left((\frac{v_0}{c})^2 - \lambda_{Di}^2 K^{2} \right) \,,
\end{equation}
where $K$ is the dimensional wave-number.  

The resemblance between (\ref{e25}) and Burger's equation suggests the use of a Cole-Hopf transformation, which converts the Burger's equation into a diffusion equation. This is a convincing method to see that the dissipative term of Burger's equation prevents the formation of a shock. Therefore, it is illustrative to consider the Cole-Hopf transformation also for (\ref{e25}), 
\begin{equation}
\label{s2}
B = \frac{2}{q}\frac{\partial q}{\partial x} \,,
\end{equation}
where $q = q(x,t)$ is a function to be determined. Inserting (\ref{s2}) into (\ref{e25}) yields
\begin{equation}
\label{s3}
\frac{\partial}{\partial x}\Bigl[\frac{\partial q/\partial t - \partial^{2}q/\partial x^2}{q} - \ln q \Bigr] = 0 \,,
\end{equation}
or
\begin{equation}
\label{s4}
\frac{\partial q}{\partial t} = \frac{\partial^2 q}{\partial x^2} + q (\ln q + \Gamma(t)) \,,
\end{equation}
where $\Gamma = \Gamma(t)$ is an arbitrary function of time only. However, this gauge function can be eliminated by 
\begin{equation}
\label{s5}
\bar{q} = e^g q \,, 
\end{equation}
where $g = g(t)$ satisfies the linear equation
\begin{equation}
\label{s6}
\dot{g} - g + \Gamma = 0 \,,
\end{equation}
so that $\bar{q}$ satisfies the same equation (\ref{s4}) as $q$, but with $\Gamma = 0$. 
Since $q$ and $\bar{q}$ produces the same magnetic field through (\ref{s2}), one can define $\Gamma \equiv 0$ without loss of generality to arrive at
\begin{equation}
\label{s7}
\frac{\partial q}{\partial t} = \frac{\partial^2 q}{\partial x^2} + q \ln q  \,.
\end{equation}
It would be a diffusion equation, except for the last (nonlinear) term. 

\begin{figure*}[ht]
\begin{center}
{\par\centering
\resizebox*{0.80\textwidth}{!}{\rotatebox{0}{\includegraphics{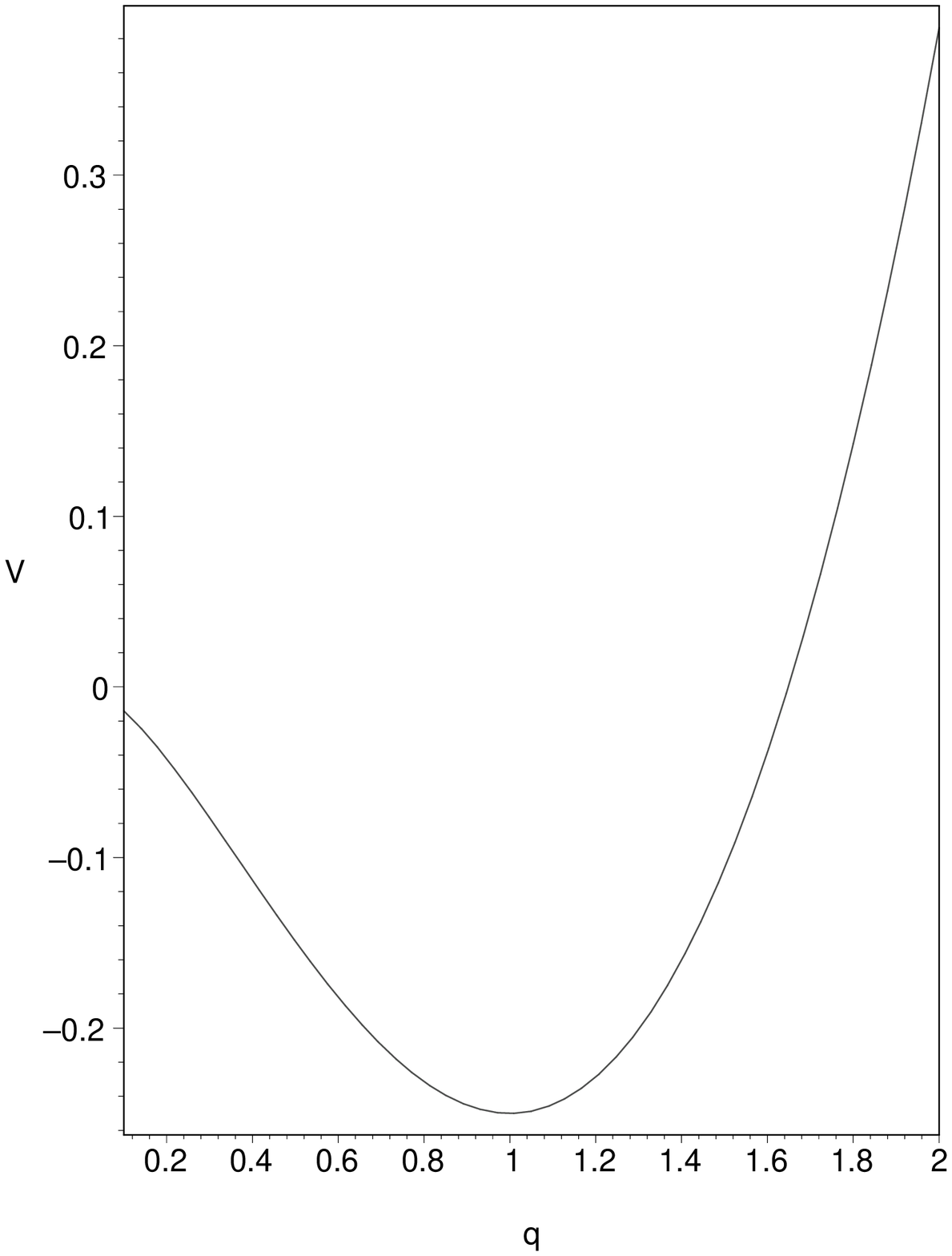}}}
\par}
\caption{Pseudo-potential $V(q)$ as defined in (\ref{s8}).}
\label{fig2}
\end{center}
\end{figure*}

The stationary states where $\partial\psi/\partial t = 0$ are associated to the solutions of a Hamiltonian system, 
\begin{equation}
\label{s8}
\frac{d^2 q}{dx^2} = - \frac{dV}{dq} \,, \quad V = \frac{q^2}{2}(\ln q - \frac{1}{2}) \,.
\end{equation}
From Figure \ref{fig2} for the potential $V$, one conclude about the existence of stable periodic solutions around $q = 1$ and hence about the periodic structure of the stationary magnetic fields. These results applies directly also to the case of a current-carrying electron-ion plasma \cite{Niknam}.

\section{Conclusion}

In this work, a perpendicular (with respect to the main flow direction) drifting ion population was considered as a possible source of periodic magnetic structures in dusty plasmas. The basic equations reduce to a partial differential equation of diffusive type for the $z$-component of the magnetic field. For small amplitude solutions, it was identified a linear instability, while the time-independent solutions are described by Lienard's equation. The periodic character of the solutions for the Lienard's equation can be associated to nonlinear magnetic field structures in space and laboratory dusty plasmas. The results are applicable, in particular, for highly collisional dusty plasmas, where there is a noticeable momentum transfer due to dust-neutrals collisions. Natural improvements are the inclusion of the ion and electron masses, as well as taking into account ion-dust or electron-dust momentum exchanges. Also a relevant open question is about the (nonlinear) stability of the predicted magnetic field structures. 

\vskip .5cm
{\bf Acknowledgments}
\vskip .5cm

This work was partially supported by the Alexander von Humboldt Foundation and by the DFG through the SFB 591. 

\vskip 1cm
\begin{center}
{\bf Appendix \\
\vskip .5cm
Conditions of validity for neglecting the Hall effect}
\end{center}
\vskip .5cm

A sensible condition for disregarding the Hall currents is given by 
\begin{equation}
\label{a1}
|v_{dx}| >> |v_{dy}| \,,
\end{equation}
where $v_{dy}$ is the $y-$component of the velocity of the dust fluid. Indeed, if (\ref{a1}) is satisfied, than the drift motion would not be an essential ingredient. 

Taking into account the $y$ component of the force equation for the dust and $E_y = - \partial A_{y}/\partial t$ as the $y-$component of the electric field, one get 
\begin{equation}
\label{a2}
v_{dy} = \frac{Z_{d}e}{m_d \nu_{dn}}\frac{\partial A_y}{\partial t} \,,
\end{equation}
where dimensional variables are used. It is evident from (\ref{a1}-\ref{a2}) that the magnetic field should not change in a too fast time-scale. While the condition is clearly achieved in equilibrium, this is not what happens in the linear regime when approaching the singularity $k \rightarrow k_{-}$, with $k_{-}$ given in (\ref{e20}). More exactly, using (\ref{e10}) and (\ref{a2}) and from now on the rescaled variables, (\ref{a1}) is rephrased as 
\begin{equation}
\label{a3}
\left|\frac{\partial A_y}{\partial t}\right| << \frac{m_d c \nu_{dn}}{Z_d m_i v_{Ti} \omega_{pi}} \left|\frac{\partial \phi}{\partial x}\right| \,.
\end{equation}
In (\ref{e3}), $v_{Ti} = (\kappa_B T_{i}/m_{i})^{1/2}$ is the ion thermal velocity and the ion plasma frequency is $\omega_{pi} = (n_{i0}e^{2}/(m_i \epsilon_{0}))^{1/2}$. Except if the electrostatic potential significantly changes in a short spatial distance, (\ref{a3}) is usually fairly satisfied  both in laboratory and astrophysical plasmas. 

More detailed conclusions can be found in the linear regime, which is also the most problematic. Indeed, diffusion and nonlinear effects are expected to prevents the development of the singularity predicted for $k \rightarrow k_{-}$. For small amplitude waves, (\ref{e15}) gives 
\begin{equation}
\label{a4}
\phi = - \frac{\partial^2 A_y}{\partial x^2} - A_y \,.
\end{equation}
Taking into account (\ref{a4}) and after Fourier transforming (\ref{a3}), there follows
\begin{equation}
\label{a5}
Im(\omega) << \frac{m_d c \nu_{dn}}{Z_d m_i v_{Ti} \omega_{pi}} k |1 - k^{2}| \,,
\end{equation}
where $\omega$ is given by the dispersion relation (\ref{e18}). Now some algebra shows that (\ref{a5}) is equivalent to $k >> k_m$, where
\begin{equation}
\label{a6}
k_m = \frac{1-\rho  + \Biggl[(1-\rho )^2 + 4 \,\frac{T_i \rho }{T_e} (\frac{m_d c \nu_{dn}}{Z_d m_i v_{Ti} \omega_{pi}})^2 (1 + \frac{T_i \rho }{T_{e}}) \Biggr]^{1/2}}{2 m_d c \nu_{dn}(1 + \frac{T_i \rho }{T_{e}})/(Z_d m_i v_{Ti} \omega_{pi})} \,.
\end{equation}

In particular, for the fully depleted electrons case, the smallest allowable wave-number reduces to the much simpler form 
\begin{equation}
\label{a7}
k_m = \frac{Z_d m_i v_{Ti} \omega_{pi}}{m_d c \nu_{dn}} \,, \quad \rho  \simeq 0 \,,
\end{equation}
while, in the absence of dust, it reduces to
\begin{equation}
\label{a8}
k_m = \left(1 + \frac{T_e}{T_i}\right)^{-1/2} \,, \quad \rho  \simeq 1 \,.
\end{equation}
For instance \cite{DAngelo}, in comet tails with $n_{e0} \simeq n_{i0} = 10 \,cm^{-3}$ and a small fraction $n_{d0}/n_{i0} \simeq 10^{-6}$, one can apply (\ref{a8}) with $T_{e}/T_i \simeq 5$, yielding $k_m = 0.41$. Also notice that in the pure electron-ion case, $k_m$ as given by (\ref{a8}) is the same as $k_{-}$ in (\ref{e19}). 

As apparent from either (\ref{a6}) or (\ref{a7}), without discussing the r\^ole of the remaining parameters, the range of validity of the model increases for a larger dust-neutrals collision frequency. This is natural, since perpendicular drifts tends to be inhibited in a strongly collisional medium. As an example, consider the plasma parameters in the vicinity of Saturn's main $A$ and $B$ rings, as recently measured by the Cassini spacecraft orbiter and shown in Table 1 of \cite{Yaroshenko}. Such planetary rings are basically a dusty plasma with low degree of ionization. Indeed, the Cassini measurements reports on an oxygen atmosphere and an ionospheric plasma near Saturn's main rings. There the ratio between the ion $O_{2}^{+}$ and neutral molecular oxygen $O_2$ densities is as low as $10^{-3} - 10^{-4}$, so that ion-neutral momentum transfer is expected to be a relevant process. For \cite{Yaroshenko} the plasma parameters $\rho \simeq 1/10, T_i \simeq T_e \simeq 1-10 \, eV, m_d = 4 \times 10^{-12} g, \nu_{dn} \simeq 10^{-11} - 10^{-10} s^{-1}, Z_d \simeq 1 - 10^2, v_{Ti}(O_{2}^{+}) \simeq 2 \times 10^5 - 7 \times 10^5 cm s^{-1}, \omega_{pi}(O_{2}^{+}) \simeq 2.5 \times 10^2 - 1.2 \times 10^3 s^{-1}$, one get from (\ref{a6}) the range $k_m \simeq 0.3 - 0.7$. Better results could be found for outer rings like Saturn's $F$ ring, where $\rho$ has much smaller values \cite{Barkan}. For these distances, however, the plasma ions significantly overtake the dust grains, so that momentum exchanges between ions and charged dust grains should be considered \cite{Yaroshenko}. In all cases, the singularity predicted by the linear theory for $k = k_{-}$ is unphysical, since $k_{-}$ as given by (\ref{e19}) is smaller than $k_m$. Considering the large degree of depletion, if one uses (\ref{g}) with $K = 0$ and $v_{0}/c = 10^{-2}$ one roughly estimate the maximum growth-rate as $0.8 - 1.9 \times 10^4 s^{-1}$. 

A practical way to reach smaller wave-numbers is to enlarge the dust-neutrals collision frequency, which can be calculated  \cite{DAngelo} from 
\begin{equation}
\label{a9}
\nu_{dn} \simeq \frac{4 m_N N a^2 v_{TN}}{m_d} \,,
\end{equation}
where $m_N$ is the neutrals mass, $N$ the neutrals number density, $v_{TN}$ the neutrals thermal speed and where the dust grains where assumed to be all spheres of radius $a$. Therefore, for space plasmas or for laboratory experiments, for smaller $k_{m}$'s one should have, in particular, the largest possible neutrals mass density combined to the largest neutrals thermal velocity. In addition, it follows from (\ref{a6}) that larger wave-lengths are achieved for larger values of $T_{e}/T_i$.  

As a final example, one can choose parameters appropriated to the dust-acoustic instability. The dust-acoustic experiments are promising, since they consider a small ionization fraction ($\simeq 10^{-5}$), so that the ion drag force is smaller than the neutral friction even if $\nu_{dn}$ is smaller than the ion-dust collision frequency \cite{Ivlev}. Following \cite{Merlino}, take $\rho = 1/2$, $T_i = 300 K$, $T_e = 23000 K$, $m_i = m_N = 4.7 \times 10^{-23} g$, $m_d = 10^{-9} g$, $Z_d = 4 \times 10^4$, $v_{Ti} = v_{TN} = 3 \times 10^4 cm s^{-1}$, $\omega_{pi} = 10^3 s^{-1}$, $N = 3 \times 10^{15} cm^{-3}$, $a = 5 \times 10^{-4} \,cm$. Combining (\ref{a6}) and (\ref{a9}), there follows $k_m = 0.08$. This small value comes basically from the large ratio $T_{e}/T_i$.  Finally, the maximal growth-rate can be estimated using (\ref{e18}), since now the depletion is not so high. Considering (\ref{e18}) with $k = k_{m}$, and for $v_{0}/c = 10^{-2}$, there follows the value $8.1 \times 10^3 \, s^{-1}$ for the growth-rate.


\begin{thebibliography}{99}
\bibitem{Yaroshenko} V. V.  Yaroshenko, F. Verheest and G. Morfill, Astron. and Astroph. {\bf 461}, 385 (2007) (and references therein).
\bibitem{Ivlev} A. V. Ivlev, D. Samsonov, J. Goree, G. Morfil and V. E. Fortov, Phys. Plasmas {\bf 6}, 741 (1999). 
\bibitem{Shukla1} P. K. Shukla and M. Rosenberg, Phys. Plasmas {\bf 6}, 1038 (1999). 
\bibitem{Barkan} A. Barkan, R. L. Merlino and N. D'Angelo, Phys. Plasmas {\bf 2}, 3563 (1995).
\bibitem{DA} N. D'Angelo, Phys. Lett. A {\bf 304}, 102 (2002). 
\bibitem{Rosenberg} M. Rosenberg, J. Vac. Sci. Technol. A {\bf 14}, 631 (1996). 
\bibitem{DAngelo} N. D'Angelo, Planet. Space Sci. {\bf 46}, 1671 (1998).
\bibitem{Merlino2} R. L. Merlino, IEEE Trans. Plasma Sci. {\bf 25}, 60 (1997). 
\bibitem{Merlino} R. L. Merlino and N. D'Angelo, Phys. Plasmas {\bf 12}, 054504 (2005).
\bibitem{Rosenberg2} M. Rosenberg, Planet. Space Sci. {\bf 41}, 229 (1993). 
\bibitem{Mizra} A. M. Mizra, M. A. Mahmood and G. Murtaza, New J. Phys. {\bf 5}, 116 (2003).
\bibitem{Avinash} K. Avinash and A. Sen, Phys. Lett. A {\bf 194}, 241 (1994). 
\bibitem{Niknam} A. R. Niknam and B. Shokri, Phys. Plasmas {\bf 15}, 012108 (2008).
\bibitem{Polyanin} A. Polyanin and V. F. Zaitsev, {\it Handbook of Exact Solutions for Ordinary Differential Equations} (Chapman-Hall/CRC, Boca Raton, 2003), p. 301. 
\bibitem{Shukla} P. K. Shukla and A. A. Mamun, {\it Introduction to Dusty Plasma Physics}, (IOP Publishing, Bristol, 2002), p. 4.
\bibitem{Mamun} A. A. Mamun, R. A. Cairns and P. K. Shukla, Phys. Plasmas {\bf 3}, 702 (1996). 
\bibitem{Farid} T. Farid, A. A. Mamun, P. K. Shukla and A. M. Mizra, Phys. Plasmas {\bf 8}, 1529 (2001). 

\end{thebibliography}
\end{document}